# Measurements of W/Z Production in Association with Jets at D0

Ashish Kumar
*The State University of New York at Buffalo, NY 14260, USA*
(on behalf of the D0 Collaboration)

Understanding the associated production of jets and vector bosons is of paramount importance for the top quark physics, for the Higgs boson and for many new physics searches. In this contribution, recent measurements of W/Z+jets and Z+b-jets processes by the D0 experiment are presented. The measurements are compared to theoretical predictions from next-to-leading order (NLO) perturbative QCD calculations where available, and to several Monte Carlo model predictions.

## 1. Introduction

Study of the associated production of vector bosons with jets constitutes an important topic in the high-$p_T$ physics program at the Tevatron. Measurements of production cross sections and kinematic distributions of W/Z+jets provide a stringent test of quantum chromodynamics (QCD) predictions at high $Q^2$ and offer the possibility to validate Monte Carlo (MC) simulation tools. Furthermore, W/Z+jets processes constitute a significant source of background in many measurements and searches at the Tevatron and the LHC, like top-quark production measurements and Higgs boson and super-symmetric particles searches.

In this contribution, a review of the recent results on vector boson plus jet production by the D0 [1] experiment is presented. Various analyses use proton-antiproton collision data from the Fermilab Tevatron at the center of mass energy √s=1.96 TeV corresponding to integrated luminosities of 1.0 fb$^{-1}$ to 4.2 fb$^{-1}$. The results are corrected for the experimental effects and presented at the "particle level". The theory predictions are corrected for the hadronization and underlying effects which are obtained from the parton shower Monte Carlo (MC) simulations.

## 2. Z/γ*+Jets Production

The D0 experiment has measured differential cross sections for Z/γ*(→e$^+$e$^-$)+jets+X production using a data sample corresponding to integrated luminosity of 1 fb$^{-1}$[2]. The measurements are binned in the $p_T$ of the N$^{th}$ jet, using events containing at least N = 1, 2, or 3 jets, and are normalized to the measured inclusive Z/γ*(→e$^+$e$^-$)+X cross section to reduce uncertainties. The Z boson is selected via its decay into a pair of high-$E_T$ electrons whose invariant mass is compatible with the mass of the Z. Jets are defined using the Run II mid-point algorithm [3] with cone size R = 0.5 and are required to satisfy $p_T$ > 20 GeV and |η| < 2.5.

In Fig. 1, the measured jet $p_T$ spectra are compared with the perturbative QCD predictions from MCFM v5.3 [4] at next-to-leading order (NLO) and leading order (LO) for the leading jet. Both the LO and NLO predictions are seen to agree with data within experimental and theoretical uncertainties over one order of magnitude in jet $p_T$ and four orders of magnitude in cross section. As expected, the NLO prediction has significantly lower scale uncertainties than the LO prediction, corresponding to a higher predictive power. In Fig.1, the measurements are also compared with the predictions of various commonly used event generators. The parton-shower based HERWIG and PYTHIA Tune QW event generator models show significant disagreements with data which increase with jet $p_T$ and the number of jets in events. The $p_T$ -ordered shower model in PYTHIA gives a good description of the leading jet, but shows no improvement over the old model for the sub-leading jets. The SHERPA and ALPGEN+PYTHIA generators show an improved description of data as compared with the parton-shower-based generators. ALPGEN+PYTHIA gives a good description of the shapes of the jet $p_T$ spectra, while predicting lower production rates than observed in data. SHERPA, on the other hand, predicts higher production rates and a less steeply falling jet $p_T$ spectrum for the leading jet than observed in data. For ALPGEN+PYTHIA, the factorization and renormalization scales can be chosen so that a good, simultaneous agreement with data is achieved for all three leading jets. For SHERPA, a similar level of agreement is achieved for the sub-leading jets, but some disagreements remain for the shape of the leading jet $p_T$ spectrum.

## 3. Angular Distributions in Z/γ*+Jets Production

The D0 experiment has performed the first measurements of the angular correlations between the Z/γ* and leading jet in Z/γ*+jet+X production using an integrated luminosity of 0.97 fb$^{-1}$ [5]. Using the decay mode Z/γ*→ μ$^+$μ$^-$, differential

Z/γ*+jet+X cross sections are measured, binned in the azimuthal angle between the Z/γ* and leading jet, |Δφ(Z, jet)|, the absolute value of the rapidity difference between the Z/γ* and leading jet, |Δy(Z, jet)|, and the absolute value of the average rapidity of the Z/γ* and leading jet, |$y_{boost}$(Z + jet)|. These differential cross sections are normalized to the measured inclusive Z/γ* cross section, cancelling many systematic uncertainties. The angular distributions are sensitive to QCD radiation. These measurements are therefore excellent tests of the inclusion of QCD radiation in theoretical models.

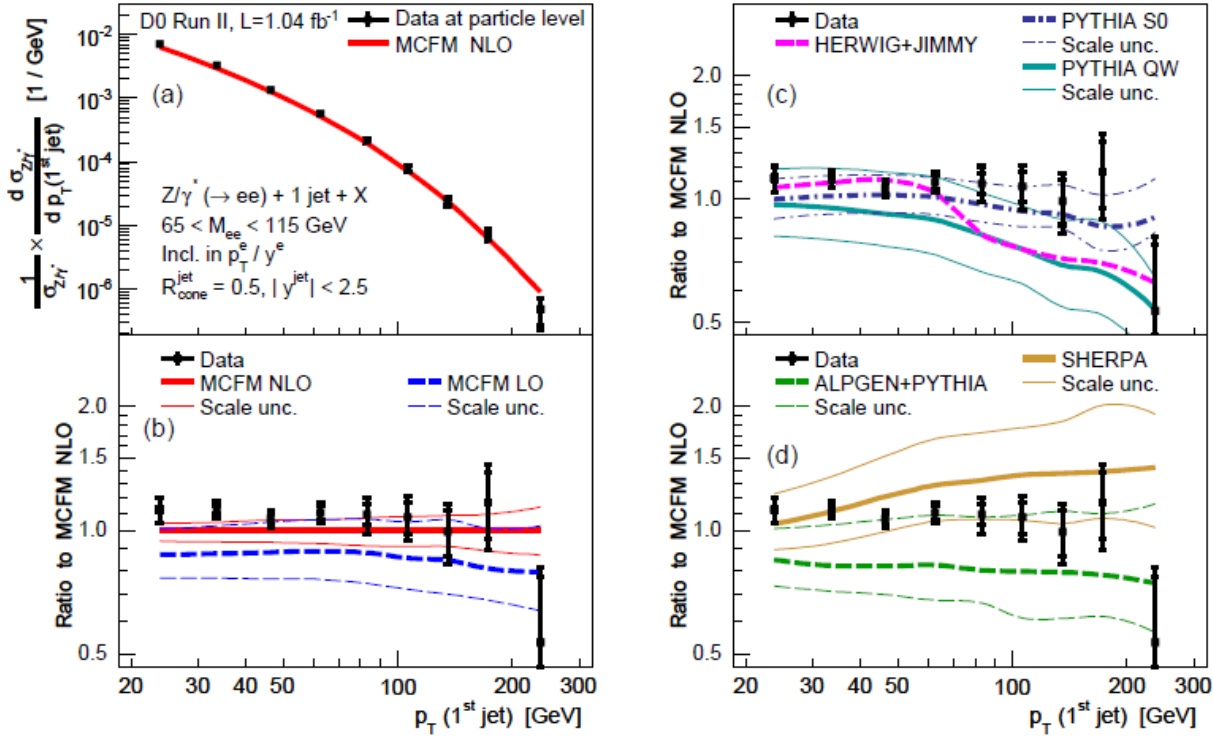

Figure 1: (a) The measured differential cross section of the leading jet in Z/γ*+jet+X events normalized to the inclusive Z/γ* cross section, compared to the predictions from MCFM NLO. The ratios of data to theory predictions from MCFM NLO are shown (b) for pQCD predictions corrected to the particle level, (c) for three parton-shower event generator models, and (d) two event generators matching matrix-elements to a parton shower MC. The scale uncertainties are evaluated by varying the factorization and renormalization scales by a factor of two.

The Z boson is selected via its decay into a pair of high-$E_T$ muons whose invariant mass is compatible with the mass of the Z boson. Jets are defined using the Run II mid-point algorithm with cone size R = 0.5 and are required to satisfy $p_T$ > 20 GeV and |η| < 2.8. The normalized differential cross section binned in Δφ(Z, jet) is presented in Fig. 2. The measured spectra are compared with LO and NLO pQCD calculations obtained using MCFM. The NLO pQCD prediction, in the Δφ(Z, jet) range where it is available, provides a reasonable description of the data and exhibits a significant improvement in both, shape and uncertainty over LO. Comparisons are also made to PYTHIA with tunes QW and Perugia, ALPGEN interfaced to PYTHIA with the same tunes, HERWIG (with JIMMY for multiple parton interactions) and SHERPA. The ratios in Fig. 2 are shown with respect to the prediction of SHERPA and the shaded band indicates the corresponding scale uncertainty. All event generators suffer from significant scale uncertainties, of comparable size to the uncertainty on LO prediction. Among the studied event generators, SHERPA provides the best description of the shape of the data, but shows a normalization difference. HERWIG shows significant disagreement with data. The modelling of Δφ(Z, jet) is improved when PYTHIA is interfaced to ALPGEN. These studies of the kinematics of inclusive Z production at D0 complement previous cross section measurements in which the boson decays into electron and muon pairs.

### 4. W+jets Production

D0 has recently presented measurements of W(→eν) + n jets cross-sections for n=1,2,3 and 4 jets using a data sample of 4.2 fb[-1] [6]. The measurements include the total inclusive cross section for each jet multiplicity and differential cross sections normalized to inclusive W cross section as a function of $n^{th}$ jet $p_T$. The measurements improve on the previous CDF results by including the W+4 jet differential cross sections, and by significantly improving the uncertainties on the differential cross sections as well as by performing the first comparison with NLO W+3 jet predictions using BLACKHAT+SHERPA [7] and ROCKET+MCFM [8]. Figure 3 shows the ratio of the predictions to the measured differential cross sections as a function of the $n^{th}$ jet $p_T$ in W+n jet events. For the W+4 jet production, only LO prediction is available. The measured cross sections are generally found to agree with the theory, although certain regions of phase space are identified where these predictions could better match the data.

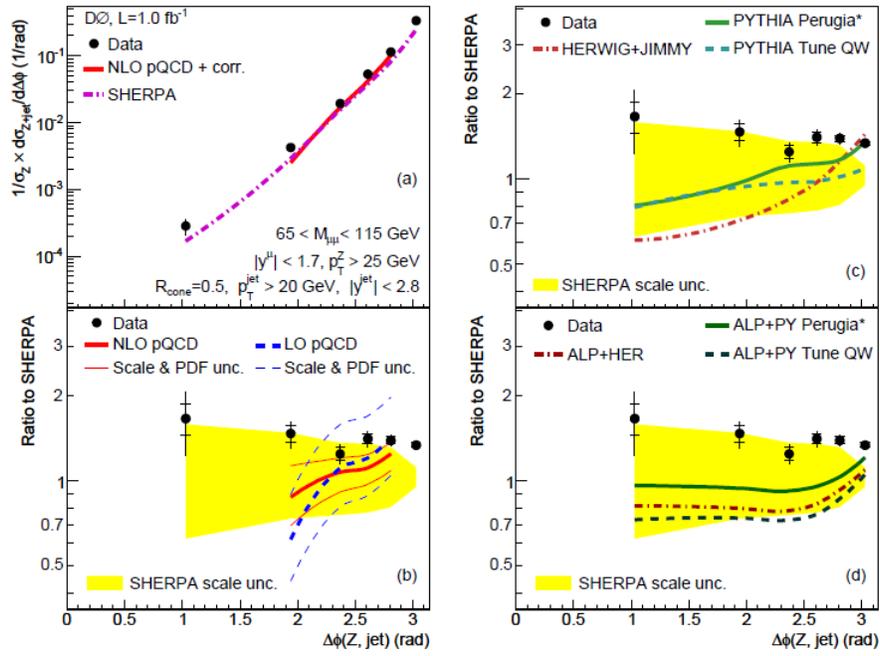

Figure 2: The measured normalized cross section in bins of $\Delta\phi(Z, jet)$ for $Z/\gamma^*$+jet+X events for Z $p_T$> 25 GeV. The distribution is shown in (a) and compared to fixed order calculations in (b), parton shower generators in (c), and the same parton shower generators matched to ALPGEN matrix elements in (d). All ratios in (b), (c), and (d) are shown relative to SHERPA, which provides the best description of data overall.

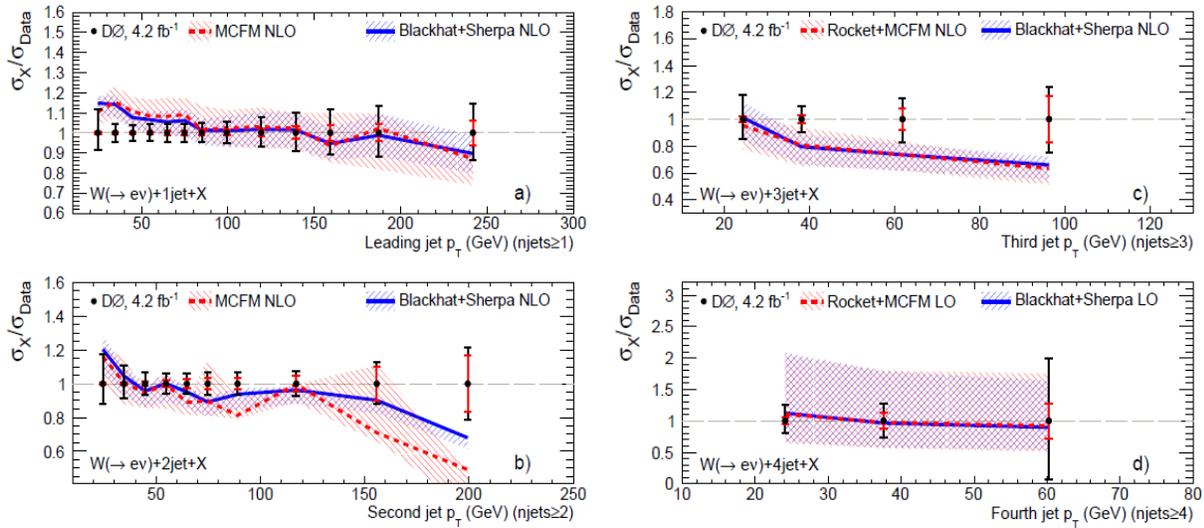

Figure 3: The ratio of the predictions to the measured differential cross sections as a function of the $n^{th}$ jet $p_T$ in (a) W+1 jet events, (b) W+2 jet events, (c) W+3 jet events, and (d) W+4 jet events.

## 5. Z + b Jet Production

The D0 experiment has published the most precise measurement of the ratio of cross-sections for the Z+b-jet to Z+jet production based on about 4.2 fb$^{-1}$ of data [9]. The measurement of the ratio benefits from cancellations of many systematic uncertainties, and therefore allows a more precise comparison with theory. Events that are consistent with the *Z* boson decays to electrons or muons that contains at least one jet (cone size R=0.5, $p_T$ > 20 GeV and $|\eta|$ < 2.5) are selected. The b-jets are identified by means of a dedicated neural network (NN) technique that exploits the various characteristics of tracks of particles inside the jet [10]. The b-jet fraction in the sample enriched with heavy flavors is extracted using a state of the art discriminant constructed from impact parameter significance of tracks associated with the jet and the secondary vertex mass. Figure 4 shows the discriminant distribution of b-tagged jets for data along with the fitted contributions from jets of different flavors. The measured ratio $\sigma(Z+b)/\sigma(Z+jet)$ = 0.0193 ±0.0022 (stat) ± 0.0015 (syst) is in good agreement with the theoretical prediction of 0.0192 ± 0.0022.

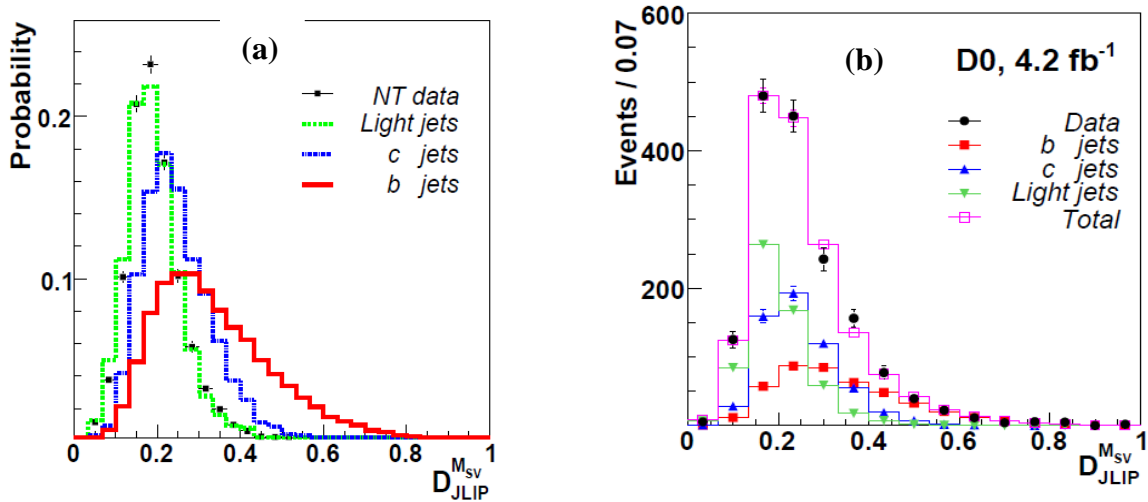

Figure 4: (a) The distributions of the discriminant for *b*, *c*, and light jets passing the b-tagging requirement which are used for fitting the data to extract the fraction of b-jet. (b) The distributions of the b, c and light jets are weighted by the fractions found from the fit to data.

## 6. Summary

Final states containing a vector boson and jets appear in many interesting physics processes. The good understanding of QCD production of such final states is critical for physics analyses at the Tevatron and LHC. The D0 measurements of Z/W + jets and Z/W + heavy flavor production in general are in good agreement with the theoretical predictions, although with large uncertainties in some cases. Since all presented measurements are fully corrected for detector effects, they can be directly used for testing and improving existing and future theory models. The measurements with improved precision from the Tevatron experiments would be the key for deeper understanding of these processes and benefit for future physics analyses at the Tevatron and LHC.